\documentclass[preprint,12pt]{elsarticle}

\usepackage{amssymb}
\usepackage[cmex10]{amsmath}
\usepackage{amsfonts}
\usepackage{array}
\usepackage{url}

\usepackage[]{fancyhdr} % 

\usepackage{caption}
\usepackage{subcaption}

\usepackage{algorithm}
\usepackage{algorithmic}

\usepackage{tablefootnote}
\usepackage{braket}
\usepackage{bm}
\usepackage{xcolor}
\usepackage{tikz}
\usetikzlibrary{quantikz}
\usepackage{circuitikz}
\usepackage[capitalise]{cleveref}
\usetikzlibrary{shapes.geometric,arrows,calc,patterns}
\tikzstyle{platform} = [rectangle, rounded corners, minimum width=10mm, minimum height=10mm,text centered, draw=black]
\tikzstyle{lab} = [rectangle, minimum width=5mm, minimum height=2mm, text centered, draw=black,fill=white]
\tikzstyle{method} = [rectangle, minimum width=5mm, minimum height=5mm, text centered, draw=black,fill=gray!30]
\tikzstyle{arrow} = [thick,->,>=stealth]
\usepackage{multicol,lipsum}
\usepackage{multirow}
\tikzstyle{startstop} = [rectangle, rounded corners, minimum width=0.5cm,text width=1.5cm, minimum height=0.5cm,text centered, draw=black]
\tikzstyle{io} = [trapezium, trapezium left angle=70, trapezium right angle=110, minimum width=2.0cm, minimum height=0.75cm, align=center, inner sep=1ex, draw=black]
\tikzstyle{process} = [rectangle, minimum width=0.5cm,text width=1.5cm, minimum height=1.5cm, text centered, draw=black]
\tikzstyle{decision} = [diamond, minimum width=1.2cm,minimum height=1.2cm,aspect=2, align=center, inner sep=-1ex,draw=black]

\makeatletter
\newcommand\notsotiny{\@setfontsize\notsotiny\@vipt\@viipt}
\makeatother

\usepackage[utf8]{inputenc}
\usepackage{pgfplots}
\DeclareUnicodeCharacter{2212}{−}
\usepgfplotslibrary{groupplots,dateplot}
\usetikzlibrary{patterns,shapes.arrows}
\pgfplotsset{compat=newest}

\tikzset{
    load/.style={
        draw,
        shape border rotate=180,
        regular polygon,
        regular polygon sides=3,
        fill=black,
        node distance=0.1cm,
        minimum height=0.1cm
    }
}

\tikzset{
    external/.style={
        draw,
        regular polygon,
        regular polygon sides=4,
        pattern=crosshatch,
        node distance=0.1cm,
        minimum height = 1.0cm
    }
}

\newsavebox\mybox
\usepackage{subcaption}
\newsavebox{\largestimage}

\journal{Electric Power Systems Research}

\begin{document}

\begin{frontmatter}



\title{Stochastic Quantum Power Flow for Risk Assessment in Power Systems}


\author{Brynjar Sævarsson, Hjörtur Jóhannsson, Spyros Chatzivasileiadis
\\ \{brysa, hjjo, spchatz\}@dtu.dk
} %% Author name

\affiliation{organization={Department of Wind and Energy Systems, \\ Technical University of Denmark},%Department and Organization
            % addressline={Elektrovej 325}, 
            city={Kgs. Lyngby},
            % postcode={2800}, 
            country={Denmark}}

%% Abstract
\begin{abstract}

This paper introduces the first quantum computing framework for Stochastic Quantum Power Flow (SQPF) analysis in power systems. The proposed method leverages quantum states to encode power flow distributions, enabling the use of Quantum Monte Carlo (QMC) sampling to efficiently assess the probability of line overloads. Our approach significantly reduces the required sample size compared to traditional Monte Carlo methods, making it particularly suited for risk assessments in scenarios involving high uncertainty, such as renewable energy integration. We validate the method on two test systems, demonstrating the computational advantage of quantum algorithms in reducing sample complexity while maintaining accuracy. This work represents a foundational step toward scalable quantum power flow analysis, with potential applications in future power system operations and planning. The results show promising computational speedups, underscoring the potential of quantum computing in addressing the increasing uncertainty in modern power grids.

\end{abstract}


% \begin{highlights}
% % These are too long. MAX 85 characters, with spaces.
% % \item Encoding power system applications in a quantum circuit can unlock the benefits of quantum algorithms
% % \item Quantum sampling from a quantum encoded line flow distribution requires fewer samples than classical Monte Carlo for same accuracy
% \item The first framework for Stochastic Quantum Power Flow is developed.
% \item Quantum algorithms can enhance the efficiency of stochastic power flow analysis.
% \item Fewer samples needed than classical Monte Carlo for accurate risk assessment.
% \item Quantum Amplitude Estimation can effectively estimate probability of line overload.
% \item Quantum Monte Carlo offers potential speedup in power system analysis.
% \end{highlights}

%% Keywords

\end{frontmatter}


\newpage
\section{Introduction}

The increasing share of renewable energy sources (RES) in power systems creates a challenge to ensure system security and optimize future system planning. Weather-based energy sources, such as wind and solar, rely on forecasts that introduce some degree of uncertainty in the system operation. At the same time, with increasing distributed generation, increased electrification, and highly varying loads such as electric vehicles, consumption patterns are also becoming more difficult to predict. This increasing uncertainty in both generation and consumption means that the number of possible operating scenarios grows exponentially, making it infeasible to evaluate them all. With systems being operated closer to their limits, classical deterministic methods based on peak scenarios do not provide the full picture of the risk factors and can lead to a false dimensioning of the system. Methods for addressing stochastic power flow have become increasingly important in ensuring the security of the power system. 

Several Stochastic Power Flow (SPF) methods have been developed using classical computations \cite{plf1974, ppfbook}. Monte Carlo simulations, based on random sampling, are commonly used to address these types of problem, but to achieve results with an acceptable confidence level, a large number of samples are required. The number of samples depends on the desired accuracy and can be in the range of tens to hundreds of thousands. For SPF, a commonly used number of power flow samples is $10\,000$ \cite{popf, pscopf}. Although this may not be computationally challenging for a single estimate, the need for better accuracy, shorter time frames, and especially when combined with a security assessment such as the N-1 security criterion, the number of required simulations quickly reaches millions, and here the computational burden can become significant. Here, the emerging technology of quantum computing has the potential to become a powerful tool for power system applications. 

Quantum Computing (QC) has been shown to achieve significant speedups for certain computational problems compared to classical computers. For power systems, active research is focused on areas where QC can speed up computations and help alleviate computational bottlenecks. Various Quantum Power Flow (QPF) applications have been explored based on the HHL quantum algorithm \cite{hhl}, which has a theoretical exponential speedup for solving a set of linear equations. The DC power flow is solved in \cite{q_dc_pf}. In \cite{q_ac_pf} the authors lay the foundation for a quantum AC power flow method and also suggest using their approach in stochastic power flow analysis based on classical Monte Carlo sampling. Expanding on the AC power flow, in our previous work \cite{sævarsson2022quantum},  we use for the first time real quantum computers to compute QPF. We also show that extracting the power flow result kills the speedup given by the HHL algorithm, making it unsuitable for this approach. This is further explored in \cite{pareek2024demystifying} where the end-to-end complexity of QPF is assessed. 

Quantum computing is inherently probabilistic, which indicates that it could be very suitable for stochastic methods such as SPF. There is also great promise in using Quantum Monte Carlo (QMC) sampling. QMC methods have been theoretically proven to achieve the same confidence level using only $\sqrt{N}$ of the classically needed samples \cite{qmcspeedup}, indicating that a quantum-based SPF might require only $100$ samples instead of the $10\,000$ samples needed by the classical Monte Carlo approach. This means that if a quantum computer and a classical computer would take the same time to compute one sample, applying Quantum Monte Carlo would lead to a 100x speedup. Of course, this may seem like a generous assumption as we can execute Classical Monte Carlo simulations in a High Performance Cluster taking advantage of multiple CPUs, while real Quantum Computers are not at the same level of maturity yet (November 2024). While the theoretical potential of QMC suggests a reduction in sample complexity, practical limitations in quantum hardware and encoding currently limit the realization of these benefits. As such, this work represents an essential proof of concept, with further optimizations required for practical applications. However, even if quantum computing hardware is slower than classical, the reduced number of samples could still yield an overall faster result. Considering the rapid developments in Quantum Computing, and the significant, theoretically proven, computational advantages, this paper takes one of the first steps to design algorithms which can exploit the quantum advantage for real power systems in the near future.

Quantum Monte Carlo is based on Quantum Amplitude Estimation (QAE) algorithms, which has shown promise in different fields such as finance applications \cite{riskassessment} and power system reliability assessment \cite{q_rel_ass}. Papers demonstrating the effectiveness of QAE usually assume that the distribution being sampled is known and then encoded in a quantum register. For stochastic power flow, this would mean running the classical power flow for multiple samples and then encoding the results into the quantum computer. This would essentially result in double sampling and would always be slower than just classical MC. A challenge in truly achieving the speedup of QAE is to have the initially unknown distribution encoded as a quantum state. Or, in our case, to fully perform the SPF within the quantum computer.

Quantum mechanics is fundamentally linear \cite{linearqc}, and so is quantum computing. This makes solving non-linear problems with QC a challenge. As mentioned before, previously proposed non-linear quantum power flow approaches require extracting the full solution from the quantum computer in each iteration, which cancels any quantum speed-ups. It is therefore necessary to reformulate the problem. Recent research indicates that approaches using mid-circuit measurements and classical feed-forward operations (often called dynamic circuits) show good promise in solving certain nonlinear equations \cite{non_linear}. This means measuring specific qubits in the circuit and then applying an operation according to the measurement result. However, this approach is still immature for more complicated calculations, such as the solution of the power flow, and, therefore, it is outside the scope of our work in this paper. Future work, and as the technology matures, shall explore the option of dynamic circuits for power flow calculations.

In this paper, we propose a framework for computing Stochastic Quantum Power Flow (SQPF). As a first step towards this framework, we use a modification of the linear DC power flow method, where, for each generator and load, a distribution of forecasted power with some degree of uncertainty is encoded into a quantum register. Next, unitary operations are performed to transform the injection distributions to a new quantum state that represents the line flow distributions. Finally, we use a QMC method to estimate the characteristics of the resulting distribution to identify the probability of overloading the lines in the system. The full quantum application is validated using a simulated quantum computer, as the gate depth of the current implementation of the quantum circuit exceeds the capabilities of today's Noisy Intermediate Scale Quantum (NISQ) hardware; still, substeps of the algorithm are also tested and demonstrated using real hardware. The results we obtained can be seen as an intermediate step towards developing a more accurate AC based SQPF using the same framework. The Python code developed for this article is publicly available in a git repository \cite{mygit}. 

This paper is organized as follows. \cref{sec:method} outlines the method developed. \cref{sec:qc} gives a brief introduction to quantum computing. \cref{sec:pf} describes the formulation of the stochastic quantum power flow. \cref{sec:sim} describes the simulation setup, the quantum hardware requirements, and provides the results of the power flow. In \cref{sec:discussion}, we discuss potential issues with the scalability of the method and its future potential. \cref{sec:conclusion} concludes.

\section{Risk assessment with SQPF}
\label{sec:method}

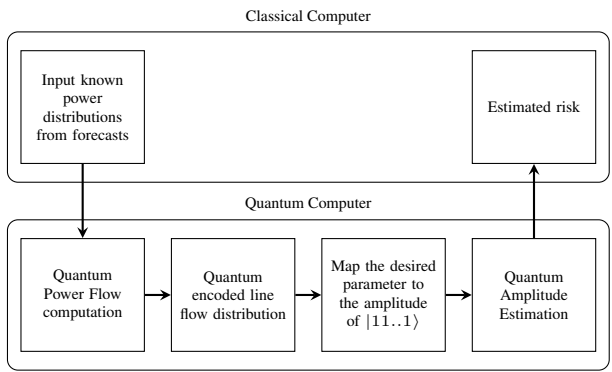
\begin{figure}[h]
\centering
\begin{tikzpicture}[node distance=2cm]
\notsotiny
\node (cc) [platform,minimum width=8cm,minimum height=2cm,fill=blue!0] {};
\node [above of=cc,yshift=-0.8cm] {Classical Computer};
\node (qc) [platform,minimum width=8cm,minimum height=2cm,fill=blue!0, below of=cc, yshift = -0.5cm] {};
\node [above of=qc,yshift=-0.8cm] {Quantum Computer};
\node (forecast) [process,xshift=-3cm] {Input known power distributions from forecasts};
\node (qpf) [process, below of=forecast, yshift = -0.5cm] {Quantum Power Flow computation};
\node (lineflow) [process, right of=qpf] {Quantum encoded line flow distribution};
\node (var) [process, right of=lineflow] {Map the desired parameter to the amplitude of $\ket{11..1}$};
\node (iqae) [process, right of=var] {Quantum Amplitude Estimation};
\node (measure) [process, above of=iqae, yshift = 0.5cm] {Estimated risk};

\draw [arrow] (forecast) -- (qpf);
\draw [arrow] (qpf) -- (lineflow);
\draw [arrow] (lineflow) -- (var);
\draw [arrow] (var) -- (iqae);
\draw [arrow] (iqae) -- (measure);

\end{tikzpicture}
\caption{Proposed framework for estimating the risk of line overload using the Stochastic Quantum Power Flow method}
\label{fig:process}
\end{figure}

The developed framework takes classically known probability distributions in a power system, e.g. a wind power forecast, with some degree of uncertainty. These distributions are then loaded into a quantum computer, which computes the power flow in the system and maps a desired feature of the result into a specific quantum state, such as the risk of line overloading. This value is then estimated using QAE and the result is extracted to a classical computer. As mentioned before, here we initially base our SQPF on the linear DC power flow but for future applications the aim is to insert an AC based power flow in the "Quantum Power Flow computation" block instead. The contribution of this paper is to introduce and successfully demonstrate the overall quantum computing framework. This already requires a substantial design process and algebraic manipulations, as we will present in this paper. Successfully introducing this framework is the main enabler for a wide range of uses in the field of power systems, including the integration of the AC power flow. The process of the developed framework is outlined in \cref{fig:process}, showing the steps of classical and quantum computations. Individual steps are described in the following sections.

\section{Quantum Computing}
\label{sec:qc}
Quantum computing is a fundamentally different type of computing which utilizes the unique properties of quantum mechanics to perform computations. This allows certain algorithms to perform exponentially faster on quantum computers compared to classical computers. Some basics of quantum computing are introduced here; for more information, the reader is directed to \cite{qcbook}. 
Quantum computers operate with quantum bits (qubits) that differ from classical bits by being able to be in a superposition of both zero and one. Quantum states are typically written using the Dirac notation, where vectors are represented using a \textbf{ket} $\ket{v}$, or a \textbf{bra} $\bra{v}$, where $\bra{v}$ is the conjugate transpose of $\ket{v}$. The state of the qubit can be represented as:

\begin{equation}
    \ket{\psi}=\alpha\ket{0}+\beta\ket{1},
    \label{eq: qubit}
\end{equation}

where the amplitudes $\alpha$ and $\beta$ are complex numbers that can be used for calculations. When the qubit is measured, it has the probability $|\alpha|^2$ of being zero and $|\beta|^2$ of being one. If another qubit is added to the system, the number of amplitudes is doubled, meaning that the information contained in a combined quantum state scales exponentially with the number of qubits. This and other special properties of qubits allow them to perform powerful computations.

Gate-based quantum computers, which we use in this study, are programmed by creating quantum circuits. These circuits are drawn with horizontal lines representing each qubit and blocks (gates) on those lines representing unitary operations on one or more qubits. The circuits are read from left to right and represent the flow of time in the evolution of the qubits, i.e. the blocks are executed sequentially until we get to the output state on the right end of the circuit. 
Each quantum gate can be represented as a unitary matrix multiplied by the quantum state of the qubit(s) it is applied to. Multiple gates can be combined to create a single unitary operation. Real quantum computers can only execute a specific set of basis gates, and this is hardware dependent. However, if the available gates form a universal set, then "ideally" they can be combined to create any unitary operation. Solving a practical problem by only applying these basis gates is a challenging task. Therefore, to develop a quantum circuit to, for example, solve a power flow problem, we take a backwards approach. We reformulate the power flow equations as a unitary matrix multiplied by a vector and then decompose the matrix into a sequence of basis gates. While this currently does not provide an optimal circuit, it makes it much simpler to develop a proof of concept. Any circuit depth mentioned in this paper is after transpiling the quantum circuit to the basis gates used by IBM hardware.
For more information on quantum circuits and the types of gates used, the reader is directed to \cite{qcbook}. 

\section{Quantum Power Flow method}
\label{sec:pf}
Since quantum computing operations are linear, we must also define our power flow approach in terms of linear operations. The linear DC power flow is therefore a good starting point for developing SQPF. In DC power flow, we can express line flows $P_{L}$ as a linear combination of bus injections $P_{B}$. This linear combination  can be represented by a constant "Power Transfer Distribution Factor" ($PTDF$) matrix which is derived from the system susceptance matrix $B_{B}$ (removing the row and column corresponding to the slack bus) and the line-to-bus susceptance matrix $B_{F}$ \cite{spyros_OPFnotes}.

\begin{align}
   &PTDF = B_{F} \cdot B_{B}^{-1} \\
   &P_{L} = PTDF \cdot P_{B} \label{eq_plpu}
\end{align}

Since line ratings vary, we scale the $PTDF$ matrix so the output of each line is in the same range. By dividing each row in the matrix with the power rating, $P_{r}$, of the corresponding transmission line, the line loading can then be calculated directly in per unit (i.e. ranging from 0 to 1) of the line rating:
\begin{equation}
   P_{L,pu} = \frac{P_{L}}{P_{r}} = \frac{PTDF}{P_{r}} \cdot P_{B} = PTDF_{r}\cdot P_{B}
\end{equation}

We assume a given wind forecast with some uncertainty is mapped into a power injection probability distribution which we use as an input to our power flow model. When bus injections are defined as probability distributions, the resulting distributions of line flows can be calculated by multiplying the $PTDF$ matrix onto each possible combination of the different values of the bus distributions. However, this can become computationally challenging when the number of buses with uncertain injection distributions becomes large, i.e. a very high-dimensional problem. Also, if the distributions are not Gaussian, the model is non-linear or the covariance of the distributions is not known, the problem is impossible to be solved with analytical methods. This is why classical Monte Carlo simulations are typically used to estimate the resulting distributions using a fixed number of random samples. However, as mentioned above, the information contained in a quantum register increases exponentially with the number of qubits, providing some interesting possibilities when working with probability distributions, and we aim to utilize this property for our application.

\subsection{Combining probability distributions with a quantum circuit}

\begin{figure}
  \centering
  \begin{subfigure}[b]{0.45\textwidth}
    \centering
    \def\rowa{0.0cm}
    \def\rowb{-1.0cm}
    \def\cola{0.0cm}
    \def\colb{0.75cm}
    \def\colc{1.5cm}
    
    \def\gs{0.7cm} % Generator size
    \def\dl{0.2cm} % Line corner distance from bus
    \def\dg{0.7cm} % Generator distance from bus
    \def\wg{0.4cm} % Generator width
    \def\bw{0.5cm} % Bus width
    \def\bt{0.1cm} % Bus thickness
    \def\lt{0.01cm} % Line thickness
    \def\lo{0.3cm} % Line offset from bus center
    \def\ls{0.35} % Load size
    \def\dld{0.3cm} % Load distance from bus
    \def\dex{0.18cm} % External grid distance from bus
    
    \begin{center}
    \begin{circuitikz}[scale=1]
    
    % BUSES
    \draw [line width=\bt] (\cola-\bw,\rowa) -- (\cola+\bw,\rowa);
    \draw [line width=\bt] (\colb-\bw,\rowb) -- (\colb+\bw,\rowb);
    \draw [line width=\bt] (\colc-\bw,\rowa) -- (\colc+\bw,\rowa);
    
    % BUS LABELS
    \draw(\cola+\lo,\rowa) node[above=-0.05cm] {\footnotesize{1}};
    \draw(\colb+\lo,\rowb) node[below=-0.05cm] {\footnotesize{3}};
    \draw(\colc+\lo,\rowa) node[above=-0.05cm] {\footnotesize{2}};
    
    % GENERATORS
    \draw (\cola,\rowa+\dg) to [/tikz/circuitikz/bipoles/length=\gs,sV,l_=\tiny$G_1$] (\cola,\rowa+\dg-\wg);
    \draw (\cola,\rowa+\dg-\wg) to[short] (\cola,\rowa);
    \draw (\colc,\rowa+\dg) to [/tikz/circuitikz/bipoles/length=\gs,sV,l_=\tiny$G_2$] (\colc,\rowa+\dg-\wg);
    \draw (\colc,\rowa+\dg-\wg) to[short] (\colc,\rowa);
    
    % LINES
    \draw [line width=\lt] (\cola+\lo,\rowa) -- (\cola+\lo,\rowa-\dl ) -| (\colc-\lo,\rowa);
    \draw [line width=\lt] (\cola-\lo,\rowa) -- (\cola-\lo,\rowa-\dl ) -- (\colb-\lo,\rowb+\dl ) -- (\colb-\lo,\rowb);
    \draw [line width=\lt] (\colc+\lo,\rowa) -- (\colc+\lo,\rowa-\dl ) -- (\colb+\lo,\rowb+\dl ) -- (\colb+\lo,\rowb);
    
    % External grid
    \node [external,shape border rotate=180,scale=\ls] at (\colb,\rowb-\dld) {};
    \draw (\colb,\rowb) to[short] (\colb,\rowb-\dex);
    
    \end{circuitikz}
    \end{center}
    
    \vspace{-5mm}
    \caption{Three bus system}
    \label{fig_three_bus}
  \end{subfigure} 
  \quad
  \begin{subfigure}[b]{0.45\textwidth}
    \centering
    \def\rowa{0.0cm}
    \def\rowb{-1.0cm}
    \def\cola{0.0cm}
    \def\colb{1.25cm}
    \def\colc{2.50cm}
    
    \def\gs{0.7cm} % Generator size
    \def\dl{0.2cm} % Line corner distance from bus
    \def\dg{0.7cm} % Generator distance from bus
    \def\wg{0.4cm} % Generator width
    \def\bw{0.5cm} % Bus width
    \def\bt{0.1cm} % Bus thickness
    \def\lt{0.01cm} % Line thickness
    \def\lo{0.3cm} % Line offset from bus center
    \def\ls{0.35} % Load size
    \def\dld{0.3cm} % Load distance from bus
    \def\dex{0.18cm} % External grid distance from bus
    
    \begin{center}
    \begin{circuitikz}[scale=1]
    
    % BUSES
    \draw [line width=\bt] (\cola-\bw,\rowa) -- (\cola+\bw,\rowa);
    \draw [line width=\bt] (\colb-\bw,\rowa) -- (\colb+\bw,\rowa);
    \draw [line width=\bt] (\colc-\bw,\rowa) -- (\colc+\bw,\rowa);
    \draw [line width=\bt] (\cola-\bw,\rowb) -- (\cola+\bw,\rowb);
    \draw [line width=\bt] (\colc-\bw,\rowb) -- (\colc+\bw,\rowb);
    
    % BUS LABELS
    \draw(\cola+\lo,\rowa) node[above=-0.05cm] {\footnotesize{2}};
    \draw(\cola-\lo-0.15cm,\rowb) node[above=-0.05cm] {\footnotesize{1}};
    \draw(\colb+\lo,\rowa) node[above=-0.05cm] {\footnotesize{3}};
    \draw(\colc+\lo,\rowa) node[above=-0.05cm] {\footnotesize{4}};
    \draw(\colc-\lo,\rowb) node[above=-0.05cm] {\footnotesize{5}};
     
    % GENERATORS
    \draw (\cola,\rowa+\dg) to [/tikz/circuitikz/bipoles/length=\gs,sV,l_=\tiny$G_1$] (\cola,\rowa+\dg-\wg);
    \draw (\cola,\rowa+\dg-\wg) to[short] (\cola,\rowa);
    
    \draw (\colc,\rowa+\dg) to [/tikz/circuitikz/bipoles/length=\gs,sV,l_=\tiny$G_2$] (\colc,\rowa+\dg-\wg);
    \draw (\colc,\rowa+\dg-\wg) to[short] (\colc,\rowa);
    
    % \draw (\cola,\rowb-\dg) to [/tikz/circuitikz/bipoles/length=\gs,sV,l_=\tiny$G_1$] (\cola,\rowb-\dg+\wg);
    \node [external,shape border rotate=180,scale=\ls] at (\cola,\rowb-\dld) {};
    \draw (\cola,\rowb-\dex) to[short] (\cola,\rowb);
    
    % LINES
    \draw [line width=\lt] (\cola-\lo,\rowa) -- (\cola-\lo,\rowb);
    \draw [line width=\lt] (\cola+\lo,\rowa) -- (\cola+\lo,\rowa-\dl ) -| (\colb-\lo,\rowa);
    \draw [line width=\lt] (\colb+\lo,\rowa) -- (\colb+\lo,\rowa-\dl ) -| (\colc-\lo,\rowa);
    \draw [line width=\lt] (\colc+\lo,\rowa) -- (\colc+\lo,\rowb);
    \draw [line width=\lt] (\cola+\lo,\rowb) -- (\cola+\lo,\rowb-\dl ) -| (\colc-\lo,\rowb);
    % \draw [line width=\lt] (\cola,\rowb) -- (\cola,\rowb+\dl+\dl/2 ) -- (\colb,\rowa-\dl ) -- (\colb,\rowa);
    \draw [line width=\lt] (\cola+\lo,\rowb) -- (\cola+\lo,\rowb+\dl ) -- (\colc,\rowa-\dl-\dl/2 ) -- (\colc,\rowa);
    
    % LOADS
    
    \node [load,shape border rotate=0,scale=\ls] at (\colb,\rowa+\dld) {};
    \draw (\colb,\rowa) to[short] (\colb,\rowa+\dld);
    % \node [load,shape border rotate=0,scale=\ls] at (\colc,\rowa+\dld) {};
    % \draw (\colc,\rowa) to[short] (\colc,\rowa+\dld);
    \node [load,shape border rotate=180,scale=\ls] at (\colc,\rowb-\dld) {};
    \draw (\colc,\rowb) to[short] (\colc,\rowb-\dld);
    % \node [load,shape border rotate=180,scale=\ls] at (\cola-\lo,\rowb-\dld) {};
    % \draw (\cola-\lo,\rowb) to[short] (\cola-\lo,\rowb-\dld);
    
    \end{circuitikz}
    \end{center}
    
    \vspace{-7mm}
    \caption{Five bus system}
    \label{fig_five_bus}
  \end{subfigure} 
  \caption{Test systems used for SQPF}
  \label{fig_systems}
\end{figure}

Here, we use the simplest possible example to explain how SQPF works. Consider a simple power system with two generators like the one shown in \cref{fig_three_bus}. Let us assume the generators are wind turbines and we have computed the expected power generation from a weather forecast with some degree of uncertainty. We define a quantum register $B_{1}$ with $N_{qb}$ qubits for bus 1. This register is used to encode the forecasted power of the generator $G_{1}$ as a probability distribution. The binary value of the quantum register is used to define the power value injected by the generator, and the amplitudes of the quantum state represent the probability of each value being generated. With $N_{qb}=2$ qubits, it is possible to represent a probability distribution with $N_{bin}=2^{N_{qb}}=4$ bins. Each bin is then assigned a power value. For simplicity, the bins are chosen to be equal to their binary value or in the range $r=[0,1,2,3]$ MW. Please note, however, the power values can be chosen arbitrarily, and with more qubits, the resolution of the values becomes exponentially higher. The quantum state of this distribution is written as follows:

\begin{equation}
    \ket{B_{1}} = b_0\ket{00} + b_1\ket{01} + b_2\ket{10} + b_3\ket{11},
\end{equation}

where $b_0$ represents the probability that the generator produces 0 MW, $b_1$ the probability that produces 1 MW, etc. 

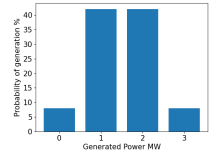
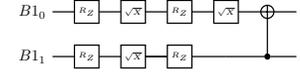
\begin{figure}[htbp]
  \centering
  \begin{subfigure}[b]{0.35\columnwidth}
    \centering
    \resizebox{0.95\textwidth}{0.7\textwidth}{%
    \begin{tikzpicture}

    \definecolor{darkgray176}{RGB}{176,176,176}
    \definecolor{darkorange25512714}{RGB}{255,127,14}
    \definecolor{lightgray204}{RGB}{204,204,204}
    \definecolor{steelblue31119180}{RGB}{31,119,180}
    
    \begin{axis}[
    legend cell align={left},
    legend style={fill opacity=1.0, draw opacity=1, text opacity=1, draw=lightgray204},
    tick align=outside,
    tick pos=left,
    x grid style={darkgray176},
    xlabel={Generated Power MW},
    xmin=-0.5, xmax=3.5,
    xtick style={color=black},
    y grid style={darkgray176},
    ylabel={Probability of generation (\%)},
    ymin=0, ymax=45,
    ytick style={color=black}
    ]
    
    \draw[draw=none,fill=steelblue31119180] (axis cs:-0.25,0) rectangle (axis cs:0.25,8.0);
    \draw[draw=none,fill=steelblue31119180] (axis cs:0.75,0) rectangle (axis cs:1.25,43.0);
    \draw[draw=none,fill=steelblue31119180] (axis cs:1.75,0) rectangle (axis cs:2.25,42.0);
    \draw[draw=none,fill=steelblue31119180] (axis cs:2.75,0) rectangle (axis cs:3.25,7.0);

    \addlegendimage{ybar,ybar legend,draw=none,fill=darkorange25512714}
    
    \end{axis}
    
    \end{tikzpicture}}
    \caption{Distribution $\ket{B_{1}}$}
  \end{subfigure}
  \quad
  \begin{subfigure}[b]{0.55\columnwidth}
    \centering
    \resizebox{.8\textwidth}{!}{%
    \begin{quantikz}
    \lstick{$B1_0$} &  \gate[1,disable auto
    height]{\text{\tiny$R_{Z}$}} & \gate[1,disable auto
    height]{\text{\tiny$\sqrt{X}$}} & \gate[1,disable auto
    height]{\text{\tiny$R_{Z}$}} & \gate[1,disable auto
    height]{\text{\tiny$\sqrt{X}$}}  &  \targ{} & \qw \\
    \lstick{$B1_1$} & \gate[1,disable auto
    height]{\text{\tiny$R_{Z}$}} & \gate[1,disable auto
    height]{\text{\tiny$\sqrt{X}$}} & \gate[1,disable auto
    height]{\text{\tiny$R_{Z}$}} \qw & \qw & \ctrl{-1} & \qw
    \end{quantikz}
    }%
    \caption{Quantum circuit for encoding $\ket{B_{1}}$}
  \end{subfigure}
  \caption{Power distribution of the generator at bus $B_1$ and the quantum circuit used for creating it}
  \label{fig_input}
\end{figure}

For an arbitrary power distribution defined for bus $i$, each element can be defined as: 

\begin{equation}
    \ket{B_{i}}_{j} = b_{j} \ket{j},
\end{equation}

where $\ket{j}$ represents the $j^{th}$ element in $r$. The combined amplitudes must satisfy that $\sum \limits_{j=0}^{N_{bin}} |b_{j}|^{2} = 1$, which means that the distribution vector, $b$, of each bus must be normalized before constructing the quantum circuit, and the final result extracted from the quantum computer must then be re-scaled with the same factor. A power distribution using two qubits is shown in \cref{fig_input} along with the quantum circuit used. The circuit consists only of the basic gates available on current quantum hardware \cite{ibmq} in order to obtain an accurate estimate of the hardware requirements for our application. The depth of the circuit is defined as the largest number of gates found in one of the rows (5 in this case).

In the same manner, a power distribution for bus 2 is encoded in a quantum register $B_{2}$. The combined quantum state, $\ket{\psi}$, then becomes the Kronecker product of the registers $B_{1}$ and $B_{2}$. 

\begin{equation}
    \ket{\psi} = \ket{B_{1}} \otimes \ket{B_{2}}
\end{equation}

This means that the quantum state $\ket{\psi}$ contains every possible combination of the two input distributions. 

\subsection{Applying $PTDFs$ with quantum computing}

For a system with $N_{l}$ lines and $N_{bus}$ buses, the resulting scaled $PTDF_r$ matrix has the shape $N_{l} \times N_{bus}$, with elements $h_{ji}$. The per unit line loading of each line can then be written as a weighted sum of the bus injections.

\begin{equation}
    L_{j} = \sum\limits_{i=0}^{N_{bus}} h_{ji} P_{B,i}
    \label{eq:Li}
\end{equation}

For a larger system, the combined quantum state of bus injections, $\ket{\psi}$ can be written as the Kronecker product of all bus distributions:

\begin{equation}
\ket{\psi}=
\bigotimes\limits_{i=0}^{N_{bus}} \ket{B_{i}}
\label{eq_bus_sv}
\end{equation}

Reformulating \eqref{eq:Li} to apply to a distribution, the resulting line loading distribution in line $j$, $ll_j$, corresponding to each amplitude in $\ket{\psi}$, given the $j^{th}$ row of the $PTDF_{r}$ matrix, can be written as a weighted Kronecker sum of the bus distributions.
 
\begin{equation}
ll_j=
\bigoplus\limits_{i=0}^{N_{bus}} h_{ji} \cdot r,
\label{eq_llj}
\end{equation}

where $\oplus$ represents the Kronecker sum of two vectors such that $A \oplus B = A \otimes e_B + B \otimes e_A$, where $e_A$ and $e_B$ are vectors of all ones with same lengths as $A$ and $B$, respectively, and $r$ is a column vector of the megawatt range defined for the amplitudes in the input registers ($r =[0,1,2,3]$ MW in our case).

We define a quantum state $\ket{L}$ whose amplitudes represent the probability of each unique value in \eqref{eq_llj}. We then define a matrix $M$, which maps the bus injection distributions, $\ket{\psi}$, to the line loading distribution $\ket{L}$ according to the structure of $ll_j$.
\begin{equation}
\ket{L} = M\ket{\psi} \label{eq:LMpsi}
\end{equation}
To construct $M$, we can define the combined probability of a certain loading in line $j$ as the sum of the quantum amplitudes that result in this loading. As with the bus distributions, we define an output range $O=0-150\%$ which encapsulates the values of the output distribution. This output range can utilize all the qubits in the circuit and therefore can have a higher number of bins than the input distributions. Combining \eqref{eq_bus_sv} and \eqref{eq_llj}, for line $j$, each amplitude in $\ket{L}$ can be defined as:
\begin{equation}
\ket{L}_{k}=\sum \left(\ket{\psi}, \: where \: ll_{j} = O_k \right)
\label{eq_ketlk}
\end{equation}
Converting \eqref{eq_ketlk} into matrix form results in $M$ being a sparse matrix with $2^{N_{bus}N_{qb}}$ columns and up to $2^{N_{bus}N_{qb}}$ rows, consisting of only zeros and ones.

\begin{figure}
    \centering
    \begin{quantikz}
    % & \text{P} & \text{M} & \text{H} & \\
    \lstick{$B1_0$} &  \gate[4,disable auto
    height]{\text{$V^{H}$}} & \gate[4,disable auto
    height]{\text{$U$}} & \qw \\
    \lstick{$B1_1$} & \qw & \qw & \qw \\ 
    \lstick{$B2_0$} & \qw & \qw & \qw \\
    \lstick{$B2_1$} & \qw & \qw & \qw \\
    \end{quantikz}
    \caption{Quantum circuit encoding $M_{sc}=UV^H$} 
    \label{fig_matcirc}
\end{figure}
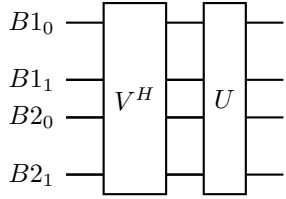

However, the matrix $M$ is not a unitary matrix. Since quantum circuits must be constructed from unitary matrices, $M$ must be decomposed into a product of unitary matrices. There are several ways of doing this, but for the type of matrix we get in \eqref{eq:LMpsi}, Singular Value Decomposition (SVD) has been found to be effective. Different decompositions were considered, but the main reason we chose SVD is that by utilizing the structure of the PTDF matrix, SVD can be used to directly decompose it as a product of two unitary matrices. Other approaches we explored require the addition of unitary matrices, which would therefore require more complex quantum circuits. SVD decomposes the matrix $M$ into two unitary matrices, $U$ and $V^H$, and a diagonal matrix, $S$, of singular values such that $M=U S V^H$. When applying SVD for an arbitrary matrix, the diagonal matrix, $S$, is usually not unitary. However, by using a simplified version of the Gram-Schmidt process, which is used to orthonormalize a set of vectors, we can get around this. Since each column in $M$ only contains a single $1$, by removing any zero rows from the matrix and normalizing the remaining rows, it can be converted to a semi-orthogonal matrix $M_{sc}$, such that $M_{sc}M_{sc}^T=I$. This ensures that the singular values of the matrix all become equal to one, which results in $S$ being an identity matrix that can be omitted in the quantum circuit. Given that $M$ consists only of zeros and ones, the $L^2$ norm for each row is the square root of the sum of columns. To achieve the correct result, this normalization factor must be reapplied later in the computation.
\begin{equation}
M_{sc,j:} = \frac{M_{j:}}{||M_{j:}||},
\end{equation}
where ${j:}$ refers to the $j^{th}$ row in the matrix. Applying SVD to the scaled matrix therefore returns two unitary matrices such that $M_{sc}=UV^H$, which can be directly plugged into a quantum circuit as in \cref{fig_matcirc}. 

An algorithm for constructing $M$ for an arbitrary line $j$ with values of row $h_{j:}$ in the $PTDFr$ matrix, and with an $N_{qb}$ power distribution at each bus is shown in \cref{alg_mat}.

\begin{algorithm}
\caption{Build Matrix}
\begin{algorithmic}
\STATE \textbf{Input} $h_{j:}, N_{bus}, N_{qb}$ 
\STATE $r = MW range$ %[0,1,2,3...2^{N_{qb}}-1]
\STATE SM = 0
\FOR{i in range($N_{bus}$)}
\STATE $W_{vec} = h_{ji}*r$
\STATE $SM = SM \oplus W_{vec}$
\ENDFOR
\FOR{k in [0,..,max($|SM|$)]}
\STATE $M[k,|SM|==k] = 1$
\ENDFOR
\RETURN $M$
\end{algorithmic}
\label{alg_mat}
\end{algorithm} 

\subsection{Extracting information from the line flow distribution}

After applying the circuit in \cref{fig_matcirc}, the line flow distribution will be encoded in a quantum state, and fully extracting the exact distribution would take a large number of runs, which would possibly kill any quantum speedups. It is therefore necessary to choose a specific aspect (i.e. "metric") of the distribution which is relevant. In this paper, we focus on the mean value of the line flow distribution and on the probability of an overloaded line.

\subsubsection{Mean value}
In quantum computing, getting the mean value of a state is commonly done by measuring the expectation value. However, due to the scaling of $M_{sc}$, the quantum encoded result is scaled. Therefore, the scaling factor of $M_{sc}$ must be reapplied to extract the correct result. Since the amplitudes of $\ket{L}$ contain the probabilities of each line loading value, given by the distinct values of the set $\{ll_{j}\}$, the mean value of the line loading can be computed as the sum of the amplitudes times the loading value they represent. This allows us to define a vector $v$ consisting of these values multiplied by the scaling factor of $M_{sc}$, which gives the re-scaled mean, $\Bar{L}$, when applied to $\ket{L}$.

\begin{align}
    &v = \{ll_{j}\} \cdot ||M_{j:}|| \label{eq:v} \\
    &\Bar{L} = v^T  \ket{L} %\label{eq_mean}
    \label{eq:Lbar}
\end{align}

To implement \eqref{eq:Lbar} in a quantum circuit, the vector $v^{T}$ must be converted to a unitary matrix. One way to do this is to apply a Househoulder transformation \cite{householder}. This transformation describes a reflection of a vector on a hyperplane. The hyperplane is defined by a unit vector $e_1$ which is orthogonal to the hyperplane. The resulting Householder matrix, $H$, is a unitary matrix and is computed as:

\begin{align}
    &\ket{\Bar{L}} = H \ket{L} \\
    &H = (I-2\frac{ww^T}{w^Tw}) \\
    &w=\frac{v}{||v||}-e_1
\end{align}

Choosing $e_1=[0,0,...,1]^T$ encodes the vector $v$ in the last row and column of $H$, resulting in the target value, $\Bar{L}$, from \eqref{eq:Lbar}, to be encoded in the amplitude of the state $\ket{1}^{\otimes n}$. 

Applying $H$ to our circuit, the final output of the power flow is stored in the quantum state $\ket{V}$.

\begin{equation}
    \ket{V} = H  M_{sc} P \ket{0} \label{eq_v}
\end{equation}

\subsubsection{Probability of line overloading}

Another interesting feature of the line flow distribution could be the probability that the line exceeds a specific load level. This can be computed using the same approach as before, using the Householder transform but with a different vector $v$. Instead of the distinct line loadings, $\{ll_{j}\}$ in \eqref{eq:v}, by setting all the values of $||M_{j:}||$ to zero except those corresponding to line load values exceeding a certain limit (4+ in the example below), these probabilities will be summed up when applied to the quantum state.

\begin{figure}
    \centering
    \tikzset{
    % operator/.append style={fill=red!20},
    % my label/.append style={above right,yshift=2.3},
    gate label/.append style={label position=above,yshift=2.0cm}
    }
    
    \tikzset{
    slice/.append style={color=black},
    % slice titles/.append style={position=below}
    }

    \begin{quantikz}
    % & \text{P} & \text{M} & \text{H} & \\
    \lstick{$B1_0$} \slice{$\ket{0}$} &  \gate[2,disable auto
    height]{\small\text{Bus injection}} \slice{$\ket{\psi}$} & \gate[4,disable auto
    height]{\small\text{Power flow}} \slice{$\ket{L}$} & \gate[4,disable auto
    height]{\small\text{Estimate}} \slice{$\ket{V}$} &  \qw  \\
    \lstick{$B1_1$}  & \qw & \qw & \qw & \qw  \\
    \lstick{$B2_0$} & \gate[2,disable auto
    height]{\small\text{Bus injection}} & \qw & \qw & \qw  \\
    \lstick{$B2_1$} & \qw & \qw & \qw & \qw
    \end{quantikz}
    \caption{Quantum circuit for computing an estimated value of a line flow distribution in the three bus system} 
    \label{fig_fullcirc}
\end{figure}
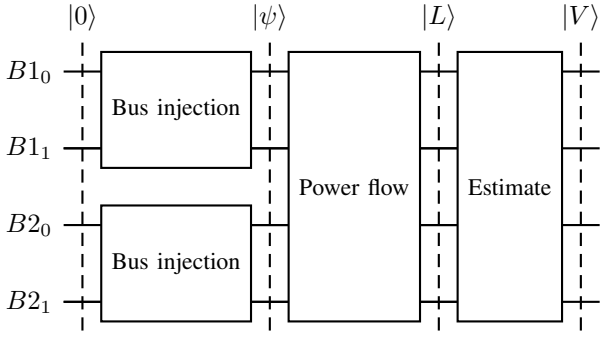

As before with the mean value, the value of interest is stored in the amplitude of $\ket{V}$ being in the one state $\ket{1}^{\otimes n}$. The combined quantum circuit constructed from \eqref{eq_v} is shown in \cref{fig_fullcirc} where the combined bus injections represent $P$, the power flow is $M_{sc}$, and the estimate is $H$. With the final value encoded in a quantum state, we now want to estimate it and extract the result. Here we apply the Quantum Monte Carlo method to estimate both cases.

\section{Quantum Monte Carlo}
\label{sec:qmc}

As mentioned before, QAE algorithms have been shown to achieve an estimate with the same confidence level using as little as $\sqrt{N}$ samples, where a classical Monte Carlo algorithm would require $N$ samples. Assuming that the quantum power flow can be executed in a time equivalent to or faster than the classical power flow methods, this would make the quantum SPF much faster than the classical SPF. Multiple algorithms have been developed to perform QAE, which differ in accuracy and hardware requirements. A comparison between some of the most common algorithms is provided in \cite{jong2023quantumamplitudeestimationprobabilistic} with a focus on their use in power systems. In this paper, we choose to apply the implementation of the Iterative QAE (IQAE) algorithm \cite{iqae} from Qiskit, which provides good accuracy and a good lower bound on the number of samples required, but has limitations for NISQ era hardware due to the required circuit depth. Among QAE methods, IQAE was chosen for its balance of accuracy and lower sampling requirements, as shown in \cite{jong2023quantumamplitudeestimationprobabilistic}. Alternative QAE algorithms like Maximum Likelihood Amplitude Estimation (MLAE) were considered but deemed less optimal. With the outlined methodology of SQPF combined with IQAE, we simulate quantum and classical computations, the results of which are discussed in the next section to highlight the efficacy of our approach. For the complete code, along with detailed representations of the quantum states at different stages of the computation and the generated full quantum circuit, the interested reader can refer to our code which we have made available online in \cite{mygit}.

\section{Simulations and results}
\label{sec:sim}
We demonstrate the Stochastic Quantum Power Flow on two test systems: a 3-bus and a 5-bus system, shown in \cref{fig_systems}. Their size is kept small so that the existing real Quantum Computers can handle their complexity. The developments in Quantum Computing show that in the near future, real QCs will be able to handle systems of at least an order of magnitude larger than the ones we use in this paper \cite{roadmap}. The power distribution of each bus, except for the slack bus, is represented by 2 qubits. As a result the 3-bus system requires 4 qubits and the 5-bus 8 qubits.

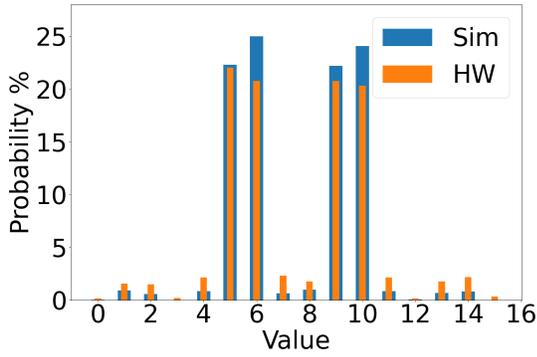
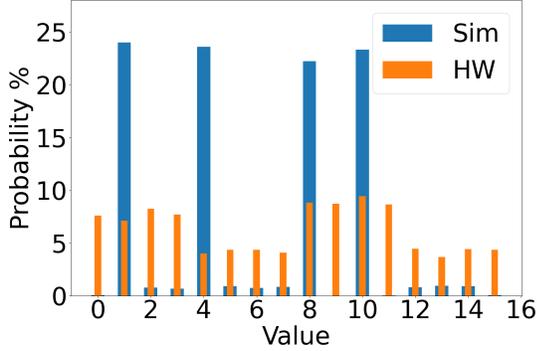
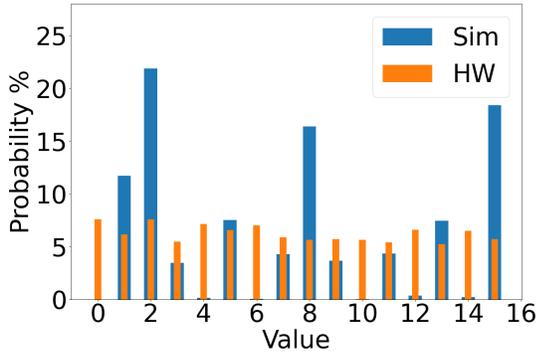
\begin{figure}[htbp]
\vspace{-50pt}
  \centering
  \begin{subfigure}[b]{0.60\textwidth}
    \centering
    \resizebox{0.95\columnwidth}{0.7\columnwidth}{%
    \begin{tikzpicture}
    
    \definecolor{darkgray176}{RGB}{176,176,176}
    \definecolor{darkorange25512714}{RGB}{255,127,14}
    \definecolor{lightgray204}{RGB}{204,204,204}
    \definecolor{steelblue31119180}{RGB}{31,119,180}
    
    \begin{axis}[
    % opacity = 0.5,
    % axis x line*= bottom,
    % axis y line*= left,
    legend cell align={left},
    legend style={fill opacity=1.0, draw opacity=1, text opacity=1, draw=lightgray204},
    tick align=outside,
    tick pos=left,
    x grid style={darkgray176},
    xlabel={Measured binary value},
    font=\footnotesize,
    xlabel style={yshift=5pt},
    xtick={0,1,2,3,4,5,6,7,8,9,10,11,12,13,14,15},
    xticklabels={0000,0001,0010,0011,0100,0101,0110,0111,1000,1001,1010,1011,1100,1101,1110,1111},
    xticklabel style={xshift=-1pt, rotate=60},
    xmin=-1.025, xmax=16.025,
    xtick style={color=black},
    y grid style={darkgray176},
    ylabel={Probability (\%)},
    ymin=0, ymax=28,
    ytick style={color=black}
    ]
    
    \addlegendimage{ybar,ybar legend,draw=none,fill=steelblue31119180}
    \addlegendentry{Sim}
    \draw[draw=none,fill=steelblue31119180] (axis cs:-0.25,0) rectangle (axis cs:0.25,0.0);
    \draw[draw=none,fill=steelblue31119180] (axis cs:0.75,0) rectangle (axis cs:1.25,0.68359375);
    \draw[draw=none,fill=steelblue31119180] (axis cs:1.75,0) rectangle (axis cs:2.25,0.68359375);
    \draw[draw=none,fill=steelblue31119180] (axis cs:2.75,0) rectangle (axis cs:3.25,0.09765625);
    \draw[draw=none,fill=steelblue31119180] (axis cs:3.75,0) rectangle (axis cs:4.25,1.5625);
    \draw[draw=none,fill=steelblue31119180] (axis cs:4.75,0) rectangle (axis cs:5.25,22.0703125);
    \draw[draw=none,fill=steelblue31119180] (axis cs:5.75,0) rectangle (axis cs:6.25,24.51171875);
    \draw[draw=none,fill=steelblue31119180] (axis cs:6.75,0) rectangle (axis cs:7.25,0.9765625);
    \draw[draw=none,fill=steelblue31119180] (axis cs:7.75,0) rectangle (axis cs:8.25,0.68359375);
    \draw[draw=none,fill=steelblue31119180] (axis cs:8.75,0) rectangle (axis cs:9.25,23.4375);
    \draw[draw=none,fill=steelblue31119180] (axis cs:9.75,0) rectangle (axis cs:10.25,22.8515625);
    \draw[draw=none,fill=steelblue31119180] (axis cs:10.75,0) rectangle (axis cs:11.25,0.5859375);
    \draw[draw=none,fill=steelblue31119180] (axis cs:11.75,0) rectangle (axis cs:12.25,0.29296875);
    \draw[draw=none,fill=steelblue31119180] (axis cs:12.75,0) rectangle (axis cs:13.25,0.9765625);
    \draw[draw=none,fill=steelblue31119180] (axis cs:13.75,0) rectangle (axis cs:14.25,0.5859375);
    \draw[draw=none,fill=steelblue31119180] (axis cs:14.75,0) rectangle (axis cs:15.25,0.0);
    
    \addlegendimage{ybar,ybar legend,draw=none,fill=darkorange25512714}
    \addlegendentry{HW}
    
    \draw[draw=none,fill=darkorange25512714] (axis cs:-0.125,0) rectangle (axis cs:0.125,0.14648438);
    \draw[draw=none,fill=darkorange25512714] (axis cs:0.875,0) rectangle (axis cs:1.125,1.5625);
    \draw[draw=none,fill=darkorange25512714] (axis cs:1.875,0) rectangle (axis cs:2.125,1.48925781);
    \draw[draw=none,fill=darkorange25512714] (axis cs:2.875,0) rectangle (axis cs:3.125,0.1953125);
    \draw[draw=none,fill=darkorange25512714] (axis cs:3.875,0) rectangle (axis cs:4.125,2.12402344);
    \draw[draw=none,fill=darkorange25512714] (axis cs:4.875,0) rectangle (axis cs:5.125,22.02148438);
    \draw[draw=none,fill=darkorange25512714] (axis cs:5.875,0) rectangle (axis cs:6.125,20.75195312);
    \draw[draw=none,fill=darkorange25512714] (axis cs:6.875,0) rectangle (axis cs:7.125,2.29492188);
    \draw[draw=none,fill=darkorange25512714] (axis cs:7.875,0) rectangle (axis cs:8.125,1.7578125);
    \draw[draw=none,fill=darkorange25512714] (axis cs:8.875,0) rectangle (axis cs:9.125,20.77636719);
    \draw[draw=none,fill=darkorange25512714] (axis cs:9.875,0) rectangle (axis cs:10.125,20.33691406);
    \draw[draw=none,fill=darkorange25512714] (axis cs:10.875,0) rectangle (axis cs:11.125,2.1484375);
    \draw[draw=none,fill=darkorange25512714] (axis cs:11.875,0) rectangle (axis cs:12.125,0.14648438);
    \draw[draw=none,fill=darkorange25512714] (axis cs:12.875,0) rectangle (axis cs:13.125,1.7578125);
    \draw[draw=none,fill=darkorange25512714] (axis cs:13.875,0) rectangle (axis cs:14.125,2.17285156);
    \draw[draw=none,fill=darkorange25512714] (axis cs:14.875,0) rectangle (axis cs:15.125,0.31738281);
    \end{axis}
    
    \end{tikzpicture}}
    \caption{Histogram when measuring $\ket{\psi}$ 1024 times, showing the combined bus injection distribution}
    \label{fig_ketpsi}
  \end{subfigure} \\
  % \quad
  \begin{subfigure}[b]{0.60\textwidth}
    \centering
    \resizebox{0.95\columnwidth}{0.7\columnwidth}{%
    \begin{tikzpicture}

    \definecolor{darkgray176}{RGB}{176,176,176}
    \definecolor{darkorange25512714}{RGB}{255,127,14}
    \definecolor{lightgray204}{RGB}{204,204,204}
    \definecolor{steelblue31119180}{RGB}{31,119,180}
    
    \begin{axis}[
    legend cell align={left},
    legend style={fill opacity=1.0, draw opacity=1, text opacity=1, draw=lightgray204},
    tick align=outside,
    tick pos=left,
    x grid style={darkgray176},
    xlabel={Measured binary value},
    font=\footnotesize,
    xlabel style={yshift=5pt},
    xtick style={color=black},
    xtick={0,1,2,3,4,5,6,7,8,9,10,11,12,13,14,15},
    xticklabels={0000,0001,0010,0011,0100,0101,0110,0111,1000,1001,1010,1011,1100,1101,1110,1111},
    xticklabel style={xshift=-1pt, rotate=60},
    xmin=-1.025, xmax=16.025,
    y grid style={darkgray176},
    ylabel={Probability (\%)},
    ymin=0, ymax=25,
    ytick style={color=black}
    ]
    
    \addlegendimage{ybar,ybar legend,draw=none,fill=steelblue31119180}
    \addlegendentry{Sim}
    \draw[draw=none,fill=steelblue31119180] (axis cs:-0.25,0) rectangle (axis cs:0.25,11.5234375);
    \draw[draw=none,fill=steelblue31119180] (axis cs:0.75,0) rectangle (axis cs:1.25,14.94140625);
    \draw[draw=none,fill=steelblue31119180] (axis cs:1.75,0) rectangle (axis cs:2.25,16.30859375);
    \draw[draw=none,fill=steelblue31119180] (axis cs:2.75,0) rectangle (axis cs:3.25,8.30078125);
    \draw[draw=none,fill=steelblue31119180] (axis cs:3.75,0) rectangle (axis cs:4.25,1.3671875);
    \draw[draw=none,fill=steelblue31119180] (axis cs:4.75,0) rectangle (axis cs:5.25,1.171875);
    \draw[draw=none,fill=steelblue31119180] (axis cs:5.75,0) rectangle (axis cs:6.25,0.09765625);
    \draw[draw=none,fill=steelblue31119180] (axis cs:6.75,0) rectangle (axis cs:7.25,0.390625);
    \draw[draw=none,fill=steelblue31119180] (axis cs:7.75,0) rectangle (axis cs:8.25,10.05859375);
    \draw[draw=none,fill=steelblue31119180] (axis cs:8.75,0) rectangle (axis cs:9.25,0.29296875);
    \draw[draw=none,fill=steelblue31119180] (axis cs:9.75,0) rectangle (axis cs:10.25,13.0859375);
    \draw[draw=none,fill=steelblue31119180] (axis cs:10.75,0) rectangle (axis cs:11.25,9.9609375);
    \draw[draw=none,fill=steelblue31119180] (axis cs:11.75,0) rectangle (axis cs:12.25,0.48828125);
    \draw[draw=none,fill=steelblue31119180] (axis cs:12.75,0) rectangle (axis cs:13.25,11.1328125);
    \draw[draw=none,fill=steelblue31119180] (axis cs:13.75,0) rectangle (axis cs:14.25,0.390625);
    \draw[draw=none,fill=steelblue31119180] (axis cs:14.75,0) rectangle (axis cs:15.25,0.48828125);
    
    \addlegendimage{ybar,ybar legend,draw=none,fill=darkorange25512714}
    \addlegendentry{HW}
    \draw[draw=none,fill=darkorange25512714] (axis cs:-0.125,0) rectangle (axis cs:0.125,7.5927734375);
    \draw[draw=none,fill=darkorange25512714] (axis cs:0.875,0) rectangle (axis cs:1.125,7.1044921875);
    \draw[draw=none,fill=darkorange25512714] (axis cs:1.875,0) rectangle (axis cs:2.125,8.251953125);
    \draw[draw=none,fill=darkorange25512714] (axis cs:2.875,0) rectangle (axis cs:3.125,7.71484375);
    \draw[draw=none,fill=darkorange25512714] (axis cs:3.875,0) rectangle (axis cs:4.125,4.00390625);
    \draw[draw=none,fill=darkorange25512714] (axis cs:4.875,0) rectangle (axis cs:5.125,4.3701171875);
    \draw[draw=none,fill=darkorange25512714] (axis cs:5.875,0) rectangle (axis cs:6.125,4.345703125);
    \draw[draw=none,fill=darkorange25512714] (axis cs:6.875,0) rectangle (axis cs:7.125,4.0771484375);
    \draw[draw=none,fill=darkorange25512714] (axis cs:7.875,0) rectangle (axis cs:8.125,8.8134765625);
    \draw[draw=none,fill=darkorange25512714] (axis cs:8.875,0) rectangle (axis cs:9.125,8.740234375);
    \draw[draw=none,fill=darkorange25512714] (axis cs:9.875,0) rectangle (axis cs:10.125,9.4482421875);
    \draw[draw=none,fill=darkorange25512714] (axis cs:10.875,0) rectangle (axis cs:11.125,8.642578125);
    \draw[draw=none,fill=darkorange25512714] (axis cs:11.875,0) rectangle (axis cs:12.125,4.443359375);
    \draw[draw=none,fill=darkorange25512714] (axis cs:12.875,0) rectangle (axis cs:13.125,3.6865234375);
    \draw[draw=none,fill=darkorange25512714] (axis cs:13.875,0) rectangle (axis cs:14.125,4.4189453125);
    \draw[draw=none,fill=darkorange25512714] (axis cs:14.875,0) rectangle (axis cs:15.125,4.345703125);
    \end{axis}
    
    \end{tikzpicture}}
    \caption{Histogram when measuring $\ket{L}$ 1024 times, showing the mapped line flow distribution}
    \label{fig_ketl}
  \end{subfigure} \\
  \begin{subfigure}[b]{0.60\textwidth}
    \centering
    \resizebox{0.95\columnwidth}{0.7\columnwidth}{%
    \begin{tikzpicture}

    \definecolor{darkgray176}{RGB}{176,176,176}
    \definecolor{darkorange25512714}{RGB}{255,127,14}
    \definecolor{lightgray204}{RGB}{204,204,204}
    \definecolor{steelblue31119180}{RGB}{31,119,180}
    
    \begin{axis}[
    legend cell align={left},
    legend style={fill opacity=1.0, draw opacity=1, text opacity=1, draw=lightgray204},
    tick align=outside,
    tick pos=left,
    x grid style={darkgray176},
    xlabel={Measured binary value},
    font=\footnotesize,
    xlabel style={yshift=5pt},
    xtick={0,1,2,3,4,5,6,7,8,9,10,11,12,13,14,15},
    xticklabels={0000,0001,0010,0011,0100,0101,0110,0111,1000,1001,1010,1011,1100,1101,1110,1111},
    xticklabel style={xshift=-1pt, rotate=60},
    xmin=-1.025, xmax=16.025,
    xtick style={color=black},
    y grid style={darkgray176},
    ylabel={Probability (\%)},
    ymin=0, ymax=25,
    ytick style={color=black}
    ]
    
    \addlegendimage{ybar,ybar legend,draw=none,fill=steelblue31119180}
    \addlegendentry{Sim}
    \draw[draw=none,fill=steelblue31119180] (axis cs:-0.25,0) rectangle (axis cs:0.25,11.71875);
    \draw[draw=none,fill=steelblue31119180] (axis cs:0.75,0) rectangle (axis cs:1.25,8.0078125);
    \draw[draw=none,fill=steelblue31119180] (axis cs:1.75,0) rectangle (axis cs:2.25,5.76171875);
    \draw[draw=none,fill=steelblue31119180] (axis cs:2.75,0) rectangle (axis cs:3.25,0.87890625);
    \draw[draw=none,fill=steelblue31119180] (axis cs:3.75,0) rectangle (axis cs:4.25,1.26953125);
    \draw[draw=none,fill=steelblue31119180] (axis cs:4.75,0) rectangle (axis cs:5.25,1.7578125);
    \draw[draw=none,fill=steelblue31119180] (axis cs:5.75,0) rectangle (axis cs:6.25,7.6171875);
    \draw[draw=none,fill=steelblue31119180] (axis cs:6.75,0) rectangle (axis cs:7.25,0.48828125);
    \draw[draw=none,fill=steelblue31119180] (axis cs:7.75,0) rectangle (axis cs:8.25,11.9140625);
    \draw[draw=none,fill=steelblue31119180] (axis cs:8.75,0) rectangle (axis cs:9.25,0.29296875);
    \draw[draw=none,fill=steelblue31119180] (axis cs:9.75,0) rectangle (axis cs:10.25,13.96484375);
    \draw[draw=none,fill=steelblue31119180] (axis cs:10.75,0) rectangle (axis cs:11.25,8.69140625);
    \draw[draw=none,fill=steelblue31119180] (axis cs:11.75,0) rectangle (axis cs:12.25,0);
    \draw[draw=none,fill=steelblue31119180] (axis cs:12.75,0) rectangle (axis cs:13.25,10.9375);
    \draw[draw=none,fill=steelblue31119180] (axis cs:13.75,0) rectangle (axis cs:14.25,0.09765625);
    \draw[draw=none,fill=steelblue31119180] (axis cs:14.75,0) rectangle (axis cs:15.25,16.6015625);
    
    \addlegendimage{ybar,ybar legend,draw=none,fill=darkorange25512714}
    \addlegendentry{HW}
    \draw[draw=none,fill=darkorange25512714] (axis cs:-0.125,0) rectangle (axis cs:0.125,7.5927734375);
    \draw[draw=none,fill=darkorange25512714] (axis cs:0.875,0) rectangle (axis cs:1.125,6.15234375);
    \draw[draw=none,fill=darkorange25512714] (axis cs:1.875,0) rectangle (axis cs:2.125,7.5927734375);
    \draw[draw=none,fill=darkorange25512714] (axis cs:2.875,0) rectangle (axis cs:3.125,5.4931640625);
    \draw[draw=none,fill=darkorange25512714] (axis cs:3.875,0) rectangle (axis cs:4.125,7.1533203125);
    \draw[draw=none,fill=darkorange25512714] (axis cs:4.875,0) rectangle (axis cs:5.125,6.591796875);
    \draw[draw=none,fill=darkorange25512714] (axis cs:5.875,0) rectangle (axis cs:6.125,7.0068359375);
    \draw[draw=none,fill=darkorange25512714] (axis cs:6.875,0) rectangle (axis cs:7.125,5.908203125);
    \draw[draw=none,fill=darkorange25512714] (axis cs:7.875,0) rectangle (axis cs:8.125,5.6396484375);
    \draw[draw=none,fill=darkorange25512714] (axis cs:8.875,0) rectangle (axis cs:9.125,5.7373046875);
    \draw[draw=none,fill=darkorange25512714] (axis cs:9.875,0) rectangle (axis cs:10.125,5.6396484375);
    \draw[draw=none,fill=darkorange25512714] (axis cs:10.875,0) rectangle (axis cs:11.125,5.419921875);
    \draw[draw=none,fill=darkorange25512714] (axis cs:11.875,0) rectangle (axis cs:12.125,6.6162109375);
    \draw[draw=none,fill=darkorange25512714] (axis cs:12.875,0) rectangle (axis cs:13.125,5.2490234375);
    \draw[draw=none,fill=darkorange25512714] (axis cs:13.875,0) rectangle (axis cs:14.125,6.494140625);
    \draw[draw=none,fill=darkorange25512714] (axis cs:14.875,0) rectangle (axis cs:15.125,5.712890625);
    \end{axis}
    
    \end{tikzpicture}}
    \caption{Histogram when measuring $\ket{V}$ 1024 times, showing the final output with the mean value stored in the amplitude of the state $1111$}
    \label{fig_ketv}
  \end{subfigure}
  \caption{Measurements at different stages in the circuit in \cref{fig_fullcirc}. The estimated states $\ket{\psi}$, $\ket{L}$ and $\ket{V}$ are shown using simulated (Sim) and real quantum hardware (HW)}
  \label{fig_measurements}
\end{figure}

The application is implemented using IBM Qiskit \cite{qiskit} and is first executed using a quantum computer simulator. In this way, we can demonstrate that the application will give accurate results on noise-free quantum computers. To compare the performance of the application in a simulated and real QC, the circuit for the three-bus system, shown in \cref{fig_fullcirc} was run in three stages. In each stage, all qubits were measured 1024 times to extract the probability distribution of the quantum states $\ket{\psi}$, $\ket{L}$ and $\ket{V}$. The resulting histograms are shown in \cref{fig_measurements}. In the first step of the circuit, that is, encoding the combined bus injections, $\ket{\psi}$, it can be seen that the real hardware result is quite similar to that of the simulator. However, due to the required circuit depth of the next steps, the noise introduced by the NISQ era hardware makes it challenging to get a satisfactory result without some form of error mitigation. Currently, a number of promising approaches are under development that bring the real hardware results close to noise-free simulations. Such methods are still under development and are, therefore, outside the scope of this paper. The primary goal of this work is to develop the first proof-of-concept and a framework for SQPF.

The total depth of the entire circuit for the 3-bus system, when transpiled for real hardware, is $863$, which is not feasible to be executed on the existing quantum computers at the moment due to its large size. But quantum computers of the very near future will be able to handle this size. For the five-bus system, the total depth is $264\,503$ and it was therefore only executed on a simulator. In \cref{fig_ketv}, the value we are looking for is encoded in the probability of measuring $\ket{1111}$ (i.e., all qubits are measured as 1), while the rest of the amplitudes are ignored, as they do not contain information relevant for our output. We can estimate the mean line loading (MLL) from the amplitude shown in the figure. 

\begin{align}
    MLL = \sqrt{Probability}*scaling \\
    scaling = ||v|| \prod_{i=1}^{N_{bus}} {||P_{i}||}
\end{align}

\begin{table}
\vspace{-20pt}
    \centering
    \begin{tabular}{c|c|c|c}
             & Probability & MLL & True value\\ \hline
         Sim & 0.1885 & 46.27478 &  \multirow{2}{*}{45.18} \\
         HW  & 0.0571 & 25.47673 &
    \end{tabular}
    \caption{Mean Line Loading (MLL) Estimate from the histogram in \cref{fig_measurements}}
    \label{tab:histresults}
\end{table}

An estimate of the mean, given the 1024 samples from \cref{fig_measurements}, can be seen in \cref{tab:histresults} where the simulated QC is quite close to the true value of $45.18$ while on real hardware the estimate is quite far off. This can be used to identify if the result is in the desired range and tells us how close the real hardware results come. However, this does not give us a metric of how accurate the solution is. In order to accurately estimate the value with a desired confidence interval, we need to apply QAE.

\subsection{QAE results}

Extracting the results through multiple measurements as shown in \cref{fig_measurements} does not provide any quantum advantage. We need to apply a QAE algorithm to the circuit to estimate the amplitude of $\ket{1111}$. We apply the standard implementation of IQAE from Qiskit, which iteratively applies the Grover operator to estimate the amplitude until the desired confidence level is achieved \cite{iqae}. Results are shown for an estimate of a single line in both three- and five-bus systems in \cref{tab:3busresults,tab:5busresults}. We estimate (i) the mean value of the line distribution and (ii) the probability that the line load exceeds $90\%$. The values estimated by IQAE clearly demonstrate the accuracy of the method. The method is set to estimate with a confidence interval $95\%$ and an error of $\epsilon=0.01$. For classical Monte Carlo, the number of required samples is computed as follows:

\begin{equation} \label{eq:NqCMC}
    N_{min}^{CMC} \sim \frac{z_c(\alpha)^2 \sigma_n^2}{\epsilon^2}, 
\end{equation}

where the critical value $z_c(\alpha)=1.96$ (for a $95\%$ confidence interval \cite{statistics}) and the normalized standard deviation is around $\sigma_n=0.754$ (found classically) in the 3 bus case and $0.722$ in the 5 bus case. For the classical approach, the margin of error in the 95\% confidence interval is given by $ME=\pm z_c(\alpha)*\frac{\sigma}{\sqrt{N}}$, where $\sigma$ is the standard deviation It can be seen that IQAE is able to estimate the value with only $12\%-16\%$ of the number of samples in both cases. The number of samples taken by IQAE and the confidence interval is returned by the algorithm and depends on how many times it has to run the circuit and the required power of the Grover operator. At this point, we shall note that recent power systems literature includes tailored approaches which can significantly reduce the required number of samples, such as \cite{MEZGHANI2020}, or approaches which can parallelize computations and reduce computation time, such as \cite{ROMANO2015}. As the goal of this paper is to introduce the first formulation for a stochastic quantum power flow, and in order to ensure a fair comparison, we compare between the Quantum and the Classical Monte Carlo. Future work will focus on scaling the Quantum Monte Carlo methods and comparing them with the tailored Classical Monte Carlo approaches.

Although the overall runtime of SQPF may currently be slower than that of the classical approach, since we need to simulate the quantum computations classically, the results in Tables~\ref{tab:3busresults} and~\ref{tab:5busresults} demonstrate a potential speedup using quantum methods under ideal conditions. In the next section, we discuss the practical implications of these findings, including potential challenges and future applications. 

\begin{table}
    \vspace{-20pt}
    \centering
    \begin{tabular}{c|c|c}
          &  Classical & IQAE  \\ \hline
         Mean line loading &  \multirow{2}{*}{45.18\% $\pm 0.45$\%} &  \multirow{2}{*}{45.19\% $\pm 0.79$\%} \\
         $\pm$ 95\% Confidence interval & & \\
         \hline
         Probability of loading over 90 \%&  \multirow{2}{*}{10.72\% $\pm 0.41$\%}&  \multirow{2}{*}{10.43\% $\pm 0.35$\%}\\
         $\pm$ 95\% Confidence interval & & \\
         \hline
         Number of samples & 21'840 & 3'500 \\
    \end{tabular}
    \caption{Results of IQAE for line 1 in the 3 bus system compared to classical results}
    \label{tab:3busresults}
\end{table}

\begin{table}
    \centering
    \begin{tabular}{c|c|c}
          & Classical &  IQAE estimate\\ \hline
         Mean line loading&  \multirow{2}{*}{33.9631\% $\pm 0.47$\%} &  \multirow{2}{*}{34.2630\% $\pm 0.65$\%}\\
          $\pm$ 95\% Confidence interval & & \\
         \hline
         Probability of loading over 90\%&  \multirow{2}{*}{2.5673\% $\pm 0.21$\%}&  \multirow{2}{*}{2.5926\% $\pm 0.50$\%}\\
          $\pm$ 95\% Confidence interval & & \\
         \hline
         Number of samples & 20'026 & 2'500 \\
    \end{tabular}
    \caption{Results of IQAE for line 1 in the 5 bus system compared to classical results}
    \label{tab:5busresults}
\end{table}

\section{Discussion}
\label{sec:discussion}

The SQPF results show that the method provides accurate results on noise-free quantum computers using almost an order of magnitude less samples than classical methods. However, the circuit depth required to run the application on real quantum computers is currently too high and scales poorly. A comparison of the computational effort between classical Monte Carlo and Quantum Monte Carlo for various stages reveals that, while quantum algorithms reduce sample count, the time to prepare and execute quantum circuits is currently substantial. For practical speedup, further circuit optimization or hardware improvement is essential.

While current quantum hardware faces challenges in executing high-depth circuits, emerging techniques, such as error mitigation and circuit decomposition, show promise in reducing noise and computational overhead. Applying error-correcting codes or segmenting computations could allow the SQPF framework to better utilize near-term quantum devices. However, even if future hardware can handle larger circuits, there is a clear need to explore ways to further reduce the required depth of the circuit to be able to efficiently scale the problem to larger power systems. If the circuit is too large, the computational overhead could negate the benefits of using QMC.
An option with which we have had success is to use quantum Shannon decomposition (QSD) \cite{shannon}. This can, for example, reduce the depth of the circuit in \cref{fig_fullcirc} by a factor of two (also for the five-bus system). This method can be applied to arbitrary unitary matrices and results in a constant number of cnot gates depending on the matrix size. For an arbitrary unitary matrix, there exists a lower bound on the number of cnot gates that scales approximately with $\mathcal{O}(4^N)$, and QSD reaches around two times the optimal. However, considering the structure of the converted PTDF matrix, it does not resemble just an arbitrary unitary matrix and likely has an even lower bound. This warrants further investigation in future work. Another approach to consider is to apply unitary approximation, where we could sacrifice some accuracy for lower circuit depth. However, this must be considered carefully, as the secure operation of power systems is highly critical and inaccurate assessment methods could lead to costly over-dimensioning of equipment or even blackouts. 

Scaling the current approach for the SQPF framework to larger systems will require substantial increases in quantum resources. For instance, the distribution loaded on each bus would require more than 2 qubits to be accurately represented. Assume we use an 8-bit resolution. Then each additional bus requires 8 qubits. This means that a 10-bus system would need at least 80 qubits. This is beyond the capabilities of currently available hardware but should be possible in the near future. Until then, focus shall be placed on continuously refining the approach to better utilize the hardware.

The approach we use to encode the probability distributions is the most straight forward for this type of application and can easily be expanded to larger systems as it uses a fixed number of qubits per bus. However, with a large number of qubits, applying the power flow matrices requires a deep circuit. Other methods we have considered, such as amplitude and angle encoding require a more complex formulation of the stochastic power flow and have therefore not been explored in as much detail. Given our experience, we would expect similar challenges with circuit depth using different encoding methods. However this is still an open question.

Currently, available methods for quantum error mitigation are only effective for small circuits. Therefore, if the SQPF circuit can be split into smaller segments, the application of these approaches could be interesting. If the noise can be sufficiently reduced, it might be possible to successfully run the application on real quantum hardware sooner. This would bring us one step closer to utilizing the theoretical advantage of quantum computers. Currently, there is a lot of research going into quantum error mitigation, and most providers of quantum systems intend to integrate these methods into their systems to provide fault tolerant quantum computations.

The linear DC power flow, which we have adapted here for quantum computations, is an important first step toward a non-linear AC SQPF. While DC power flow can be used to quickly get an approximate estimate of the system conditions, it makes several assumptions which might not make it accurate enough in highly stressed scenarios. Expanding this research to the AC Stochastic Quantum Power Flow will provide a huge advantage for future power system risk assessment. Non-linear quantum applications could also allow for a multitude of other power system assessment methods, which might be able to achieve a quantum advantage. At the same time, the approach proposed in this paper paves the way for preparing any kind of distribution that has to undergo a linear transformation before applying the QMC sampling. This could be applied to various types of analysis within power systems as Monte Carlo methods are widely used, and the quadradic speedup of QMC can provide huge benefits if the problem can be efficiently encoded as a quantum state.

Developing quantum applications for power systems requires us to reformulate our classical approaches to take advantage of the unique capabilities of quantum computers. This can be a challenging task, as quantum computers do not provide speed-ups for every type of problem. As a general comment, if we want to exploit the numerous advantages that QC can offer for power systems, we must start already now with developing the right approaches focusing for the moment on areas where quantum computers excel and see how we can utilize them for power system applications. Quantum Monte Carlo is only one of such algorithms.

\section{Conclusion}
\label{sec:conclusion}

This paper introduces, to the best of our knowledge, the first formulation for a Stochastic Quantum Power Flow (SQPF) with Quantum Monte Carlo sampling, marking a first effort to utilize the probabilistic nature of quantum computing for stochastic power flow analysis. Our approach computes power flow from an uncertain infeed and encodes the resulting distributions into quantum states, enabling the estimation of the risk for line overloads with significantly fewer samples compared to classical methods. 
We show that the Quantum Monte Carlo techniques applied together with Quantum DC Power Flow result in accurate computations while requiring an order of magnitude fewer samples than Classical Monte Carlo methods. In our work, we tested our algorithms on both real and simulated Quantum Computers. Our tests in the simulated Quantum Computers (QCs) were successful, but the real QCs still suffer from noisy hardware. With the error mitigation techniques, which are currently under development,  we expect not only that real QCs will arrive at the same level of performance as the (noise-free) simulated QCs, but also that the benefit will be greater for power systems larger than the 3-bus and 5-bus systems we tested in this paper. Future work shall focus on (i) a deeper analysis of the impact the quantum noise has on the results, as soon as the quantum hardware is able to handle quantum circuits of larger sizes, and (ii) extracting the theoretical bounds of the proposed methods which will act as a future benchmark for the tests on real hardware. Finally, intensive efforts shall continue towards the integration of the AC Power Flow in this framework, which can offer a more accurate representation of the power system. 
Continued research is essential to optimize circuit depth and scalability, ultimately unlocking the full potential of quantum computing for comprehensive and efficient power system analysis.

% Declaration of generative AI and AI-assisted technologies in the writing process.
During the preparation of this work the author(s) used Writefull for Overleaf in order to improve language and readability. After using this tool/service, the author(s) reviewed and edited the content as needed and take(s) full responsibility for the content of the publication.
\bibliographystyle{IEEEtran}
% \bibliographystyle{spmpsci}
% \bibliography{reference}
% \newpage
\bibliography{refs1}

% %% For numbered reference style

%   \textit{\LaTeX: a document preparation system},
%   Addison Wesley, Massachusetts,

% \end{thebibliography}
\end{document}